\documentclass{emulateapj}
\usepackage{color}

\shorttitle{The Origin of Long Secondary Periods in Red Giants}
\shortauthors{Soszy{\'n}ski \& Udalski}

\begin{document}

\title{The Light Curve Shapes as a Key to Resolving the Origin of Long Secondary Periods in Red Giant Stars}

\author{I. Soszy{\'n}ski\altaffilmark{1} and A. Udalski\altaffilmark{1}}
\affil{$^1$Warsaw University Observatory, Al.~Ujazdowskie~4, 00-478~Warszawa, Poland\\soszynsk@astrouw.edu.pl, udalski@astrouw.edu.pl}

\begin{abstract}
We present a study of OGLE light curves of red giant stars exhibiting long
secondary periods (LSPs) -- an enigmatic phenomenon commonly observed in
stars on the upper red giant branch and asymptotic giant branch. We show
that the light curves of LSP stars are essentially identical to those of
the spotted variables with one dark spot on their photospheres. Such a
behavior can be explained by a presence of a dusty cloud orbiting the red
giant together with a low-mass companion in a close, circular orbit. We
argue that the binary scenario is in agreement with most of the
observational properties of LSP variables, including non-sinusoidal shapes
of their radial velocity curves.
\end{abstract}


\section{Introduction}

Long secondary periods (LSPs), observed in at least one third of pulsating
red giants and supergiants, are one of the most interesting unsolved
problems of modern stellar astrophysics. This phenomenon has been known for
decades \citep{oconnell1933,payne1954,houk1963} and it is observed in tens
of thousands of long period variables (LPVs) in our and other galaxies. The
photometric amplitudes associated with the LSPs are in some cases quite
large (up to 1 mag in the {\it V} band), and radial velocity changes during
LSP cycles have been detected, but still we cannot even answer the question
of whether we are dealing with intrinsic or extrinsic stellar
variability. There are two popular hypotheses explaining the LSP
phenomenon. The first one assumes that a red giant star has a low-mass
companion which, due to interactions with the circumstellar matter, causes
the periodic photometric and spectroscopic variations
\citep{wood1999,soszynski2007}. The second hypothesis assumes that red
giant stars exhibit some kind of radial or non-radial pulsations
\citep{hinkle2002,wood2004,derekas2006}, but currently there is no
theoretical pulsation model that satisfactorily explains all the observed
features of the LSP variables.

LSPs range from about 200 to 1500 days and are an order of magnitude longer
than the typical pulsation periods of semiregular variables (SRVs) and OGLE
small amplitude red giants (OSARGs). It is worth noting that LSPs are not
observed in Mira variables, which have the largest-amplitude light curves
among pulsating red giants. In the period--luminosity (PL) diagram, LSPs
form a well-defined sequence \citep[sequence D;][]{wood1999}, roughly
parallel to other PL sequences populated by pulsating red
giants. \citet{soszynski2004} noticed that sequence D partly overlaps and
is a direct continuation of sequence E, which is formed by close binary
systems (eclipsing and ellipsoidal variables) containing a red giant as one
of the components. Sequences D and E overlap in all the studied photometric
bandpasses, from visual to infrared \citep{derekas2006}, which is a strong
argument for the binary explanation of the LSPs.

On the other hand, the shapes of the radial velocity curves are interpreted
as evidence against the binary scenario. The full velocity amplitudes
associated with LSPs range from 2 to 7~km~s$^{-1}$, with a tight clustering
around 3.5 km~s$^{-1}$ \citep{nicholls2009}. Assuming a binary origin of
the LSP phenomenon, this velocity amplitude would correspond to a secondary
component of brown-dwarf mass. The velocity curves are non-sinusoidal,
which may be interpreted as a sign of eccentric orbits. However, most of
the observed velocity curves have shapes very similar to each other, which
implies similar angles of periastron in the majority of the observed LSP
variables. Of course, one may expect that the angles of periastron in
randomly selected systems should have a uniform distribution. According to
\citet{nicholls2009}, the probability that the observed radial velocity
curves are consistent with the uniform distribution of angles of periastron
is of the order of $10^{-3}$, which practically excludes the possibility
that the LSP phenomenon is caused by binarity.

Spectroscopic observations gave us information about the changes of
the effective temperature during the LSP cycle. The changes are very small,
much smaller than expected for radial pulsation \citep{wood2004}. In
turn, nonradial oscillations would be difficult to reconcile with the
observed movement of the visible surface of the giant stars during their
LSP cycle, which is of the order of 30\% of the stellar radius
\citep{nicholls2009}. Another important observational fact was discovered
by \citet{wood2009b} -- LSP variables are surrounded by significant amounts
of cool dust and this circumstellar matter has a non-spherical (clumpy or
disk-like) distribution.

In this work we argue that the asymmetric radial velocity curve in an LSP
variable may be produced by a low-mass companion in a circular orbit just
above the surface of the red giant. Such a companion may be followed by a
dusty cloud that regularly obscures the giant and causes the apparent
luminosity variations. We present a simple model that, at least
qualitatively, well reproduces the light and velocity curves observed for
LSP stars.

\begin{figure}
\epsscale{1.1}
\plotone{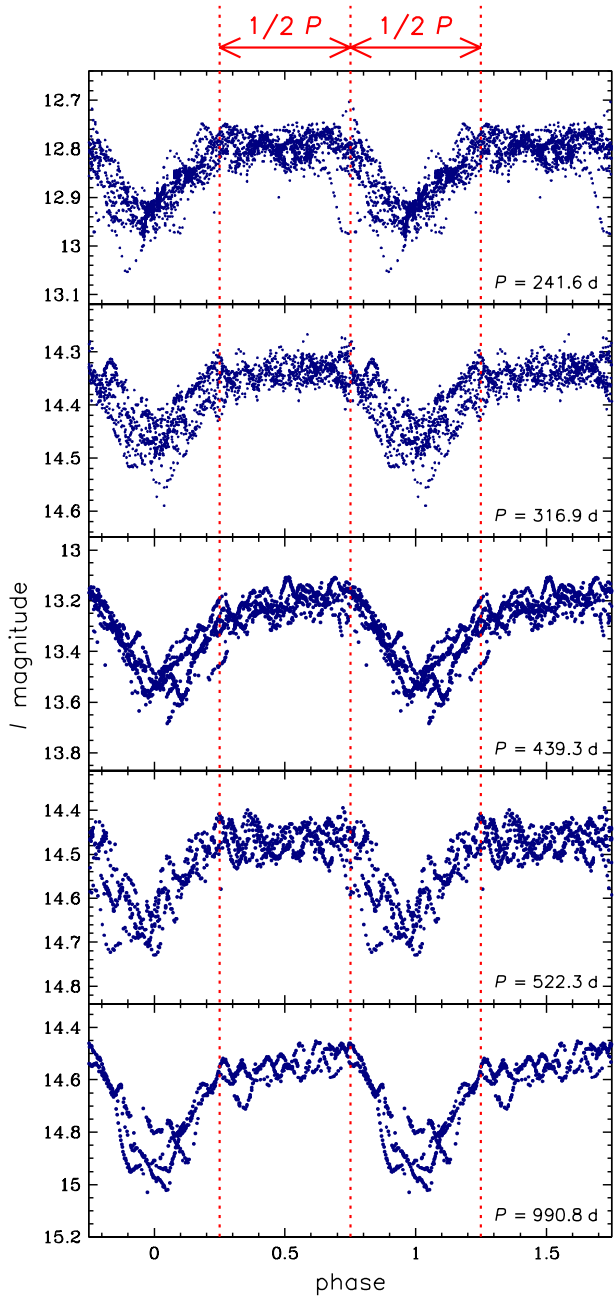}
\caption{Examples of the OGLE {\it I}-band light curves of LSP variables in
  the LMC and Galactic bulge. The light curves are folded with the LSPs
  given in the panels. Dotted vertical lines divide each light curve into
  two equal parts lasting half of the LSP cycle. See the electronic edition
  of the Journal for a color version of this figure.}
\label{fig1}
\end{figure}

\begin{figure}
\epsscale{1.1}
\plotone{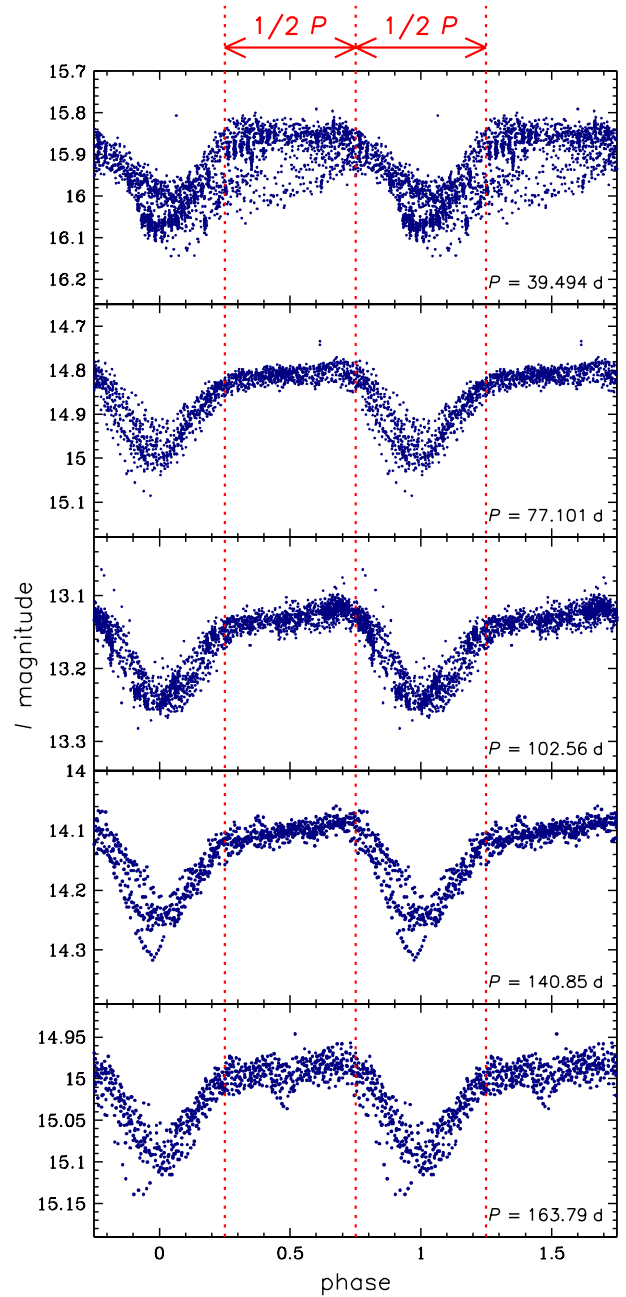}
\caption{Examples of the OGLE {\it I}-band light curves of variable stars
  with a dark spots on their photospheres. Dotted vertical lines divide
  each light curve into two equal parts lasting half of the period. See the
  electronic edition of the Journal for a color version of this figure.}
\label{fig2}
\end{figure}

\vspace{3mm}
\section{Light curves of LSP stars}

The light curves presented in this paper were obtained in the standard {\it
I}-band by the third phase of the Optical Gravitational Lensing
Experiment \citep[OGLE-III;][]{udalski2003}. The OGLE-III observations
provided 8~years (2001-2009) of continuous, high-quality light curves of
the large samples of LPVs in the Magellanic Clouds
\citep{soszynski2009,soszynski2011} and in the Galactic bulge
\citep{soszynski2013}. In this paper we analyze the photometric
observations of giants in the Large Magellanic Cloud (LMC) and Galactic
bulge.

\begin{figure*}
\epsscale{1.0}
\plotone{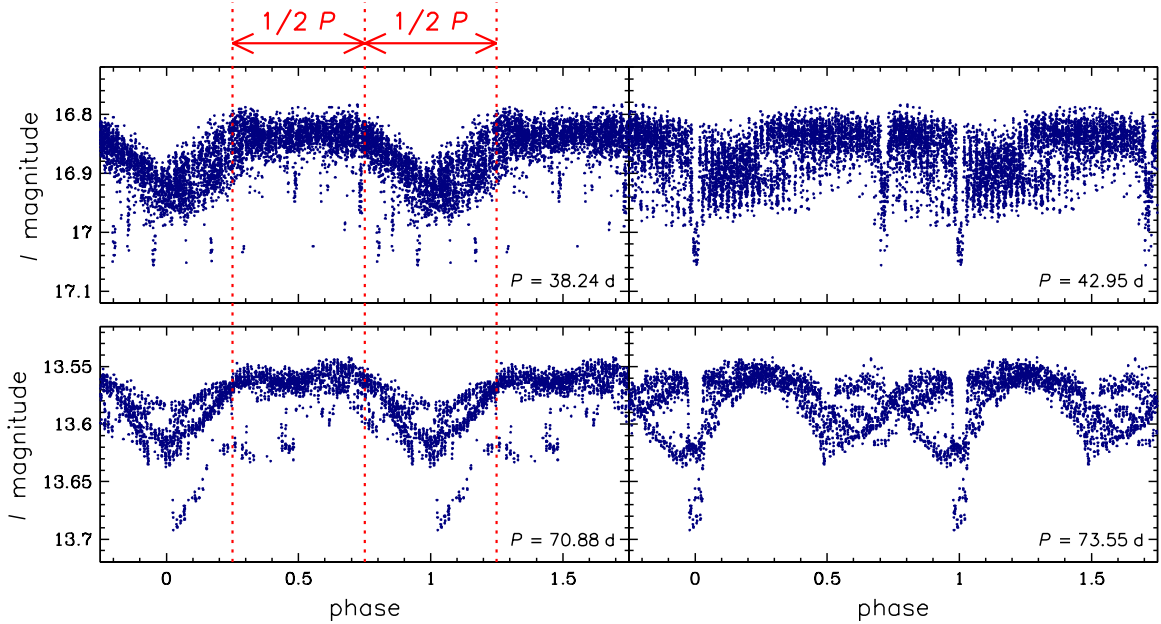}
\caption{OGLE light curves of two eclipsing RS Canum Venaticorum stars in
  the Galactic bulge. Left and right panels show the same light curves
  folded with different (although similar) periods: the periods of the
  distortion waves (left panels) and the orbital periods (right
  panels). Dotted vertical lines divide distortion waves into two equal
  parts.}
\label{fig3}
\end{figure*}

The morphology of the light curves is a characteristic that clearly
distinguishes LSP variations from the pulsations that populate PL sequences
A$'$ to C, and from the ellipsoidal variables falling on sequence~E. In
Fig.~\ref{fig1} we present several {\it I}-band light curves of LSP
variables from the LMC and bulge. These objects were selected because of
their relatively small pulsation amplitudes and quite stable amplitudes of
the LSP variations, which allowed us to highlight the typical shape of the
LSP light curves. In some sequence D stars the minima of the LSP light
curves significantly vary in depth from cycle to cycle, and even in some
cases the variations become completely invisible for a period of time
\citep{soszynski2007}.

Nevertheless, most of the LSP variables, at least those with larger
amplitudes, exhibit light curves similar to those presented in
Fig.~\ref{fig1}. As can be seen, the typical light curve of an LSP
variable may be divided into two halves. During one half of the LSP cycle,
the brightness of the star does not change at all or it increases slowly
with time. During the other half of the LSP, the light curves usually have
a triangular shape -- the luminosity decreases, reaches a minimum and then
increases. Such light curves to some extent resemble detached eclipsing
binary systems with orbital periods twice the LSPs. However, this would be
incompatible with the spectroscopic observations which show that the
velocity curves of sequence D stars vary with periods equal to the LSPs,
not twice the LSPs \citep{nie2010,nicholls2010}. Also, a comparison of the
LSP light curves with the light curves of real eclipsing binaries
\citep[e.g. collected in the OGLE samples:][]{graczyk2011,pawlak2013}
excludes the possibility that sequence D stars are classical eclipsing
binary systems (i.e. that they are eclipsed directly by the companion). The
``eclipses'' are too wide and practically in all cases there are no
alternations of deeper and shallower minima that could be interpreted as an
effect of the different temperatures of the two stellar components of the
system.

In the OGLE photometric database we found a number of red stars,
likely asymptotic giant branch stars, that have similar light curves to
the LSP variables, but have shorter periods. A sample of such light curves
is shown in Fig.~\ref{fig2}. These light curves also can be divided into
two, roughly equal parts -- one with nearly constant, sometimes slowly
increasing brightness, and the second with a symmetric decrease and
increase of brightness.

In some cases, light curves of this type additionally show narrow eclipses,
indicating that these objects are binary systems. The orbital periods of
these systems are similar, but slightly different to the periods of the
LSP-like modulation seen in their light curves. Examples of two such stars
are presented in Fig.~\ref{fig3}. Here, each light curve is shown twice: in
the left panels it is folded with the LSP-like period and in the right
panel -- with the orbital period. Objects presented in Fig.~\ref{fig3} are
typical RS Canum Venaticorum variables -- binary stars with enormous
starspots on their surfaces causing the photometric variations (the
so-called distortion waves) with periods slightly different to the orbital
periods. The difference between both periods is caused by a differential
rotation of the star which slowly changes the position of the starspots
relative to the companion. Given the similarity of the light curves shown
in Fig.~\ref{fig2} and the left panels of Fig.~\ref{fig3}, we interpret
such photometric behavior as a result of presence of starspots on the
stellar surface, namely single, relatively large, dark spots. When the spot
is hidden on the far side of a star, the apparent brightness is nearly
constant. When the spot is located on the visible side of a star, we
observe the wide minimum, which lasts roughly half of the rotation period.

\begin{figure}
\epsscale{1.17} \plotone{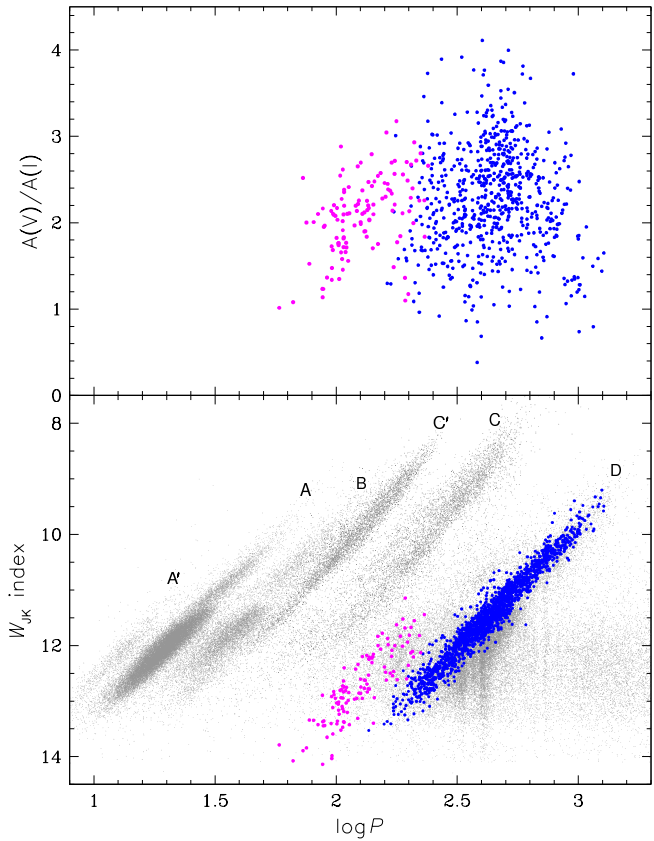}
\caption{Blue-to-red amplitude ratios as a function of period (upper panel)
  and period--Wesenheit index diagram (lower panel) for red giants in the
  LMC. Blue points indicate a selected sample of well-defined LSP
  variables, magenta points show probable spotted variables with one dark
  spot, grey points in the lower panel indicate all the LPVs in the LMC
  from the OGLE catalog \citep{soszynski2009}. [See the electronic edition
  of the Journal for a color version of this figure.]}
\label{fig4}
\end{figure}

The morphology of the light curves is not the only observational feature
that links spotted variables and LSP stars. Another such characteristic is
the ratio of light curve amplitudes measured in blue and red
filters. \citet{derekas2006} studied MACHO light curves of several thousand
LPVs and ellipsoidal variables (sequence E stars). In particular they
checked the ratios of amplitudes measured for individual objects in the
blue and red filters. They noticed that for the ellipsoidal variables the
amplitudes in both filters are very similar, which is in agreement with
expectations, since ellipsoidal variability is mainly caused by geometric
effects. LSP stars have blue-to-red amplitude ratios distinctly larger than
ellipsoidal variables and similar to the ratios of amplitudes observed in
the pulsating LPVs. For this reason, \citet{derekas2006} argued for
pulsations rather than binarity as the origin of the LSP phenomenon.

In the upper panel of Fig.~\ref{fig4} we plot the ratios of amplitudes
measured within the OGLE survey in the {\it V} and {\it I} bands against
the logarithm of periods. Here we present only those stars that have at
least 50 {\it V}-band points in the OGLE photometric database. Note that
the amplitude ratios are on average very similar for the LSP and spotted
variables with median values of $A(V)/A(I)$ equal to 2.2 and 2.1,
respectively. The conclusion is that large blue-to-red amplitude ratios may
occur not only in the pulsating stars, but also in the spotted variables
and probably in other types of stellar variability. Sequence D stars
exhibit large $A(V)/A(I)$ values which is one of the many features that
distinguish them from ellipsoidal variables \citep{nicholls2010}.

Another feature that shows similarity between spotted variables and
sequence D stars are the PL relations. The LSP variables follow the PL
relation (sequence D) which is narrowest in the near-infrared photometric
domain, especially if we use the extinction-free Wesenheit index, defined
as $W_{JK}=K-0.686(J-K)$, where $J$ and $K$ magnitudes are from the IRSF
Point Source Catalog \citep{kato2007}. In the lower panel of
Fig.~\ref{fig4} we present the PL diagram for LSP variables in the LMC
overplotted on the PL relations formed by all other LPVs detected in the
LMC by the OGLE project \citep{soszynski2009}. For clarity, we present here
only a sample of LSP variables, with well-defined light curves and
relatively large amplitudes. On the same diagram we plotted objects
classified as spotted variables with one dark spot (with light curves
similar to those shown in Fig.~\ref{fig2}). Note the PL relationship that
is followed by these variables. Periods of the spotted variables are about
two times shorter than the periods of sequence D stars of the same
luminosity.

The PL relation for spotted red giants can be explained by their fast
rotation, close to the break-up velocity, that may limit the relation from
the short-period side. The long periods of the PL relation may be limited
by the stellar magnetic activity, which is strong in fast rotators and
decreases with the increasing rotation periods. Taking this into
account, it would be difficult to explain LSPs by the presence of dark
spots in the stellar photospheres. The periods of the sequence D stars are
longer than the longest periods of spotted variables of the same
luminosity. Moreover, spectropolarimetric observations of two nearby LSP
stars performed by \citet{wood2009} did not reveal any signatures of
magnetic fields. Finally, as noted by \citet{wood2004}, the starspot
scenario would be difficult to reconcile with the observed radial velocity
curves of sequence D stars, not only with their amplitudes, but also with
the shapes of the velocity curves. A dark spot is visible roughly for a
half of the rotation period of a star, so during the other half of the LSP
cycle the velocity curve should be flat. It does not agree with the
observations which show continuous changes of the radial velocities during
the LSP cycle \citep{hinkle2002,wood2004,nicholls2009}.

\vspace{3mm}
\section{A model for the LSP variations}

It has been shown above that red giants exhibiting the LSP modulation have
similar photometric properties to spotted variables with one dark spot on
their photospheres. However, other observational facts argue against
the starspots as an explanation of the LSP phenomenon. Given the
\citet{wood2009b} discovery that sequence D stars are surrounded by a dust
in a non-spherical configuration, we propose a different explanation for
the morphology of the LSP light and velocity curves. The same
photometric effect can be produced by a dusty cloud orbiting the giant just
above its surface and regularly obscuring the star. Such a hypothesis
explaining the LSP phenomenon was proposed by \citet{wood1999} and
developed by \citet{soszynski2007}. In this scenario, the red giant has a
low-mass companion in the brown-dwarf range. The matter lost by the red
giant due to stellar winds follows a spiral pattern with a comet-like
tail. Hydrodynamical simulations of a wind-driven accretion flow in binary
systems \citep{theuns1993,mastrodemos1998,nagae2004} show that the maximum
density of such a dusty cloud is located just behind the secondary
component.

\begin{figure}
\epsscale{1.17} \plotone{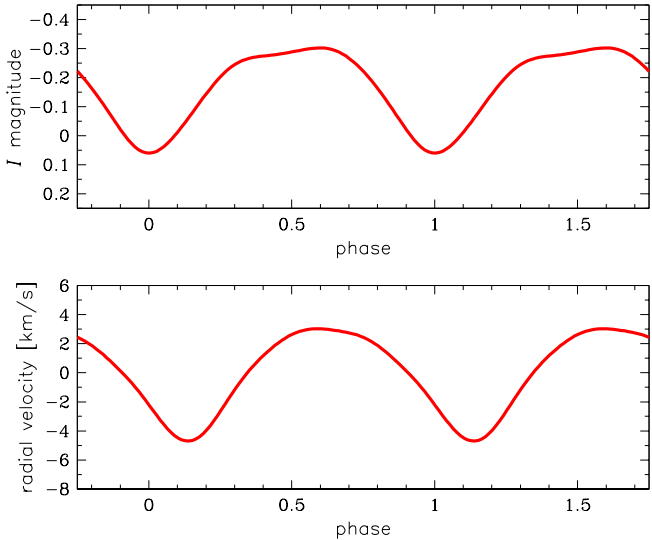}
\caption{Simulated {\it I}-band light curve (upper panel) and radial
  velocity curve (lower panel) of a red giant with a dusty cloud with a
  comet-like tail orbiting the star. The inclination of the system is
  $60^\circ$. The model was calculated with the Nightfall code.}
\label{fig5}
\end{figure}

The binary hypothesis was ruled out by \citet{wood2004} and
\citet{nicholls2009} who carried out comprehensive studies of possible
mechanisms for the LSP. Within the binary scenario, the non-sinusoidal
velocity curves of the LSP variables could be a result of eccentric
orbits. The problem with this explanation is that most of the observed
velocity curves have similar shapes, which implies similar angles of
periastron in these systems, which is inconsistent with the expectation of
random orbital orientations. This argues against binarity as the cause for
LSPs.

However, this argument may lose its validity, if we assume that the
asymmetric radial velocity curves are caused by factors other than
eccentricity of the orbits. Even if the orbit is circular, a non-sinusoidal
velocity curve may be produced from the non-spherical shape of the giant
caused by tidal interactions with its low-mass companion, and from the
dusty cloud which in different phases of its orbital cycle may block part
of the light in the approaching or declining parts of the stellar disk (the
Rossiter-McLaughlin effect). These effects added to the sinusoidal velocity
curve produced by motion in the circular orbit may give the observed
variations of the radial velocity.

We made a simple model of such a binary system with the Nightfall
code\footnote{http://www.hs.uni-hamburg.de/DE/Ins/Per/Wichmann/
Nightfall.html}, which is designed to model eclipsing binary systems and
calculate their synthetic light and velocity curves. We defined a system
with a circular orbit and mass ratio $q=0.1$. The primary star is a red
giant with the Roche lobe fill factor equal to 0.9 and co-rotating with the
secondary component, which is small and cool and practically does not
contribute to the light emitted by the system. To simulate the cloud with a
comet-like tail which periodically obscures the red giant, we defined two
dark spots on the stellar photosphere -- one located just behind the line
connecting both components (the densest part of the cloud) and the second
located about $70^\circ$ from this line (the comet-like tail of the
cloud). We tested several different positions, sizes and transparencies of
the spots to reproduce the observed features of the LSP stars.

In Fig.~\ref{fig5} we present the simulated light and velocity curves
produced by our model. It is immediately obvious that the synthetic curves
are very similar to the observed ones. The first spot (the densest part of
the cloud) causes the triangular minimum in the light curve which lasts
roughly half of the LSP. The second spot (the tail) is responsible for the
slow increase in brightness when the main part of the cloud is hidden on
the other side of the star. Despite the assumption of the circular orbit,
the simulated radial velocity curve is non-sinusoidal and resembles those
observed for LSP variables \citep{hinkle2002,wood2004,nicholls2009}.

The only inconsistency of our model with the observations is the phase
shift between light and velocity curves. In most LSP stars the minimum of
the radial velocity precedes the minimum light, although the phase shifts
vary from star to star \citep[cf. Fig.~3 in][]{nicholls2009}. In our model
the velocity lags the light by roughly 0.1 in phase. To obtain the velocity
curve that precedes the light curve, the dark spots (dusty cloud) should be
located at the opposite side of the red giant, i.e. at the longitudes more
than $180^\circ$ from the line connecting the giant and its low-mass
companion. It is not obvious if the dusty, comet-like cloud may obscure the
stellar surface at these longitudes. Clearly, more advanced modeling and
hydrodynamical simulations are needed to solve this problem.

\begin{figure}[t]
\epsscale{1.17}
\plotone{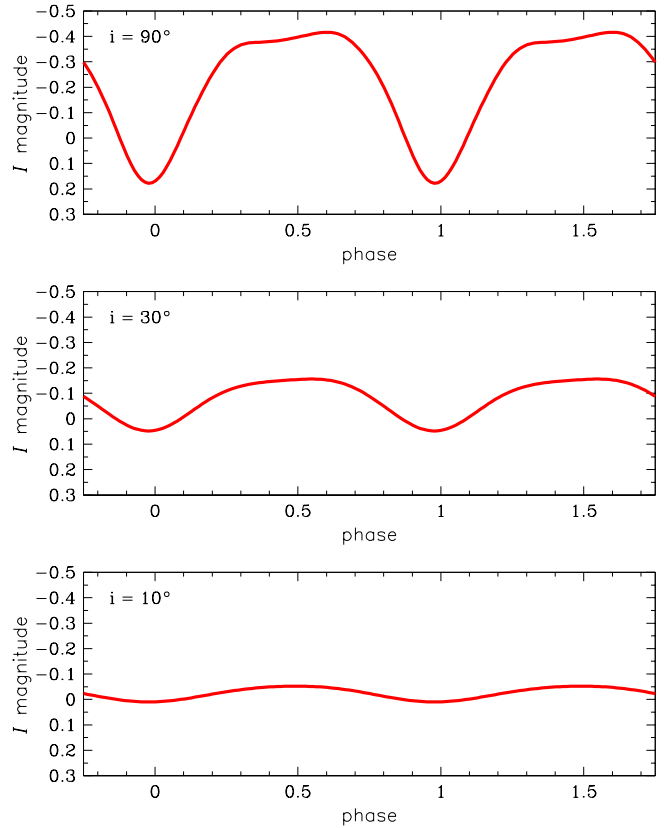}
\caption{Simulated {\it I}-band light curves of the same system as shown in
Fig.~5, but viewed at different orbital inclinations: $90^\circ$ in the
upper panel, $30^\circ$ in the middle panel, $10^\circ$ in the lower panel.}
\label{fig6}
\end{figure}

\begin{figure}[!t]
\epsscale{1.17}
\plotone{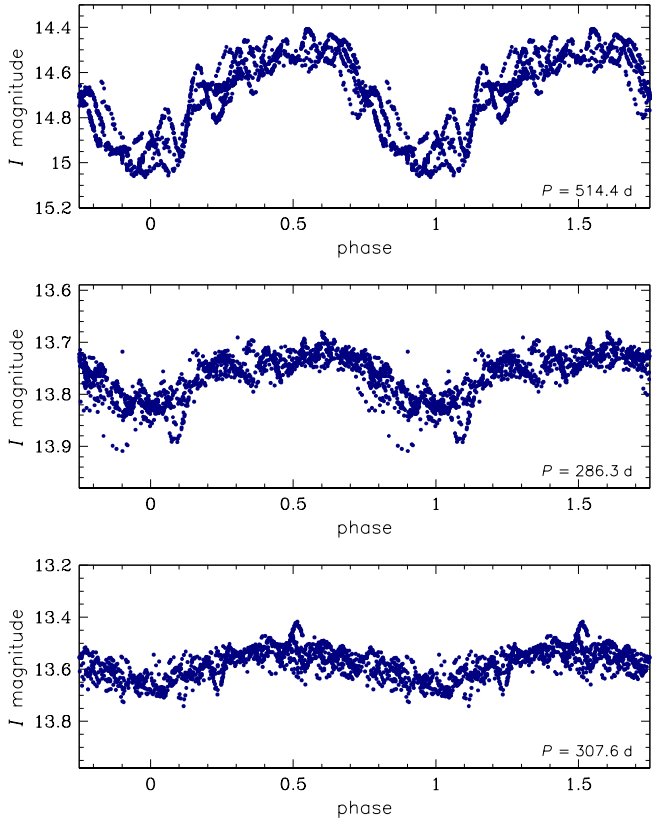}
\caption{OGLE {\it I}-band light curves of three LSP stars from the
  Galactic bulge. Note their similarity to the simulated light curves shown
  in Fig.~6.}
\label{fig7}
\end{figure}

Different sets of physical and observational parameters of the binary model
may result in different variants of the LSP light curves. For example, the
orbital inclination influences the amplitude and shape of the light
curve. In Fig.~\ref{fig6} we show three synthetic light curves of the same
system observed with the inclinations $90^\circ$, $30^\circ$ and
$10^\circ$, while in Fig.~\ref{fig7} we present light curves of three real
LSP variables with very similar morphology. To obtain other kinds of LSP
light curves, for example those which have a shorter duration of the flat
maximum, one should assume a longer and denser tail of the dusty cloud
surrounding the red giant.

\vspace{3mm}
\section{Summary and Conclusions}

Obviously, the problem of the LSP origin deserves a much more comprehensive
analysis than our simple simulation, but it seems that the binary scenario
is the only one that is consistent with virtually all the observational
facts known for LSPs. A model with a spiral dusty cloud that follows a
low-mass object on a circular orbit around the red giant well reproduces
the light and radial velocity variations associated with the LSP
modulation. This hypothesis naturally explains the observed large movements
of the visible surface of the LSP star \citep{nicholls2009}, the
mid-infrared excess caused by the circumstellar dust around LSP variables
\citep{wood2009}, the lack of significant temperature changes during the
LSP cycle \citep{wood2004}, and the PL relation of the LSPs (sequence D)
that is a direct continuation of sequence E formed by close binary systems
\citep{soszynski2004}. Large blue-to-red amplitude ratios
\citep{derekas2006} just reflect the extinction law produced by dusty
material obscuring the giant and depend on the physical properties of this
matter. Changes of the amplitudes of the LSP modulation observed in some
sequence D stars can be explained by variations of the mass-loss rates,
which influence the size and density of the dusty cloud orbiting the red
giant. The only unclear feature in our model is the reversed phase
shift between the light and velocity curves. This inconsistency can be
explained by a specific distribution of the circumstellar matter, but more
detailed models are needed to address this problem.

A natural question that arises is the origin of the brown dwarfs in the
close orbits of at least 30\% of red giant stars. These low-mass objects
could be former Jupiter-like planets that accreted part of the matter
ejected from the giants due to stellar winds, and increased their masses to
the brown dwarf range \citep{retter2005,soszynski2007}. If this explanation
were true, the LSP giants would be excellent probes of the fraction of
planets in different regions of our and other galaxies.

Another issue related to the binary explanation of the LSP phenomenon is
the mechanism that keeps the low-mass object near the giant surface for a
long time and prevents it being swallowed it by the giant companion. A
brief discussion of this problem can be found in \citet{soszynski2007b}. We
do not have a definitive answer for this question, but we suspect that the
low-mass object induces increased mass loss from the giant when its
distance is small, which in turn causes the increase of the distance
between both components. This feedback mechanism might keep a low-mass
companion in a close orbit around a giant for a long time, until the SRV
becomes a Mira star, and the changes of the giant radius due to pulsation
become very large. Then, the secondary component sinks below the giant
surface and it is completely engulfed. Therefore we do not observe LSP
modulation in the Mira stars.

As a byproduct of this study, we distinguished a group of red giants with
periodic variations which we interpret as caused by a dark spot on their
surface. These objects also follow a PL relation, spreading between
sequence~C (populated by the fundamental-mode pulsators -- Miras and SRVs)
and sequence~D (LSP). Note that \citet{soszynski2013b} noticed a dim PL
sequence located between sequences C and D for somewhat brighter giants. It
is not clear if this sequence has something in common with the spotted
variables identified in this work. Undoubtedly, both types of stellar
activity -- starspots and LSPs -- deserve special attention in future
studies.

\section*{Acknowledgments}
We are grateful to Wojciech A. Dziembowski, Peter R. Wood and Christine
P. Nicholls for carefully reading the manuscript and suggesting important
corrections. We thank the anonymous referee, whose insightful comments led
to several improvements of the paper. This work has been supported by the
Polish National Science Centre grant No. DEC-2011/03/B/ST9/02573. We
gratefully acknowledge financial support from the Polish Ministry of
Science and Higher Education through the program ``Ideas Plus'' award
No. IdP2012 000162. The research leading to these results has received
funding from the European Research Council under the European Community's
Seventh Framework Programme (FP7/2007-2013)/ERC grant agreement
no. 246678.

\label{lastpage}


\begin{thebibliography}{99}
\bibitem[Derekas et al.(2006)]{derekas2006} Derekas, A., Kiss, L.~L., Bedding, T.~R., Kjeldsen, H., Lah, P., \& Szab{\'o}, G.~M.\ 2006, \apjl, 650, L55
\bibitem[Graczyk et al.(2011)]{graczyk2011} Graczyk, D., Soszy{\'n}ski, I., Poleski, R., et al.\ 2011, Acta Astron., 61, 103
\bibitem[Hinkle et al.(2002)]{hinkle2002} Hinkle, K.~H., Lebzelter, T., Joyce, R.~R., \& Fekel, F.~C.\ 2002, \aj, 123, 1002
\bibitem[Houk(1963)]{houk1963} Houk, N.\ 1963, \aj, 68, 253
\bibitem[Kato et al.(2007)]{kato2007} Kato, D., Nagashima, C., Nagayama, T., et al.\ 2007, \pasj, 59, 615
\bibitem[Mastrodemos \& Morris(1998)]{mastrodemos1998} Mastrodemos, N., \& Morris, M.\ 1998, \apj, 497, 303
\bibitem[Nagae et al.(2004)]{nagae2004} Nagae, T., Oka, K., Matsuda, T., Fujiwara, H., Hachisu, I., \& Boffin, H.~M.~J.\ 2004, \aap, 419, 335
\bibitem[Nicholls et al.(2009)]{nicholls2009} Nicholls, C.~P., Wood, P.~R., Cioni, M.-R.~L., \& Soszy{\'n}ski, I.\ 2009, \mnras, 399, 2063
\bibitem[Nicholls et al.(2010)]{nicholls2010} Nicholls, C.~P., Wood, P.~R., \& Cioni, M.-R.~L.\ 2010, \mnras, 405, 1770
\bibitem[Nie et al.(2010)]{nie2010} Nie, J.~D., Zhang, X.~B., \& Jiang, B.~W.\ 2010, \aj, 139, 1909
\bibitem[O'Connell(1933)]{oconnell1933} O'Connell, D.~J.~K.\ 1933, Harvard College Observatory Bulletin, 893, 19
\bibitem[Payne-Gaposhkin(1954)]{payne1954} Payne-Gaposhkin, C.\ 1954, Harvard Annals, 113, No. 4
\bibitem[Pawlak et al.(2013)]{pawlak2013} Pawlak, M., Graczyk, D., Soszy{\'n}ski, I., et al.\ 2013, Acta Astron., 63, 323
\bibitem[Retter(2005)]{retter2005} Retter, A.\ 2005, Bulletin of the American Astronomical Society, 37, 1487
\bibitem[Soszy{\'n}ski(2007)]{soszynski2007} Soszy{\'n}ski, I.\ 2007, \apj, 660, 1486
\bibitem[Soszy{\'n}ski et al.(2004)]{soszynski2004} Soszy{\'n}ski, I., Udalski, A., Kubiak, M., et al.\ 2004, Acta Astron., 54, 347
\bibitem[Soszy{\'n}ski et al.(2007)]{soszynski2007b} Soszy{\'n}ski, I., Dziembowski, W.~A., Udalski, A., et al.\ 2007, Acta Astron., 57, 201
\bibitem[Soszy{\'n}ski et al.(2009)]{soszynski2009} Soszy{\'n}ski, I., Udalski, A., Szyma{\'n}ski, M.~K., et al.\ 2009, Acta Astron., 59, 239
\bibitem[Soszy{\'n}ski et al.(2011)]{soszynski2011} Soszy{\'n}ski, I., Udalski, A., Szyma{\'n}ski, M.~K., et al.\ 2011, Acta Astron., 61, 217
\bibitem[Soszy{\'n}ski et al.(2013)]{soszynski2013} Soszy{\'n}ski, I., Udalski, A., Szyma{\'n}ski, M.~K., et al.\ 2013, Acta Astron., 63, 21
\bibitem[Soszy{\'n}ski \& Wood(2013)]{soszynski2013b} Soszy{\'n}ski, I., \& Wood, P.~R.\ 2013, \apj, 763, 103
\bibitem[Theuns \& Jorissen(1993)]{theuns1993} Theuns, T., \& Jorissen, A.\ 1993, \mnras, 265, 946
\bibitem[\protect\citeauthoryear{Udalski}{2003}]{udalski2003} Udalski, A.\ 2003, Acta Astron., 53, 291
\bibitem[\protect\citeauthoryear{Wood et al.}{1999}]{wood1999} Wood, P.~R., Alcock, C., Allsman, R.~A., et al.\ 1999, in IAU Symp. 191, Asymptotic Giant Branch Stars, ed. T. Le Bertre, A. Lebre, \& C. Waelkens (Cambridge: Cambridge Univ. Press), 151
\bibitem[Wood et al.(2004)]{wood2004} Wood, P.~R., Olivier, E.~A., \& Kawaler, S.~D.\ 2004, \apj, 604, 800
\bibitem[Wood et al.(2009)]{wood2009} Wood, P. R., Marsden, S., Waite, I., \& Nicholls,~C.~P.\ 2009, in AIP Conf. Proc. 1170, Stellar Pulsation: Challenges for Theory and Observation, ed. J.~A.~Guzik \& P.~Bradley (Melville, NY: AIP), 173
\bibitem[Wood \& Nicholls(2009)]{wood2009b} Wood, P.~R., \& Nicholls, C.~P.\ 2009, \apj, 707, 573
\end{thebibliography}
\end{document}